\documentclass[prd,preprintnumbers,nofootinbib,aps]{revtex4}
\usepackage{graphicx}
\usepackage{dcolumn}
\usepackage{bm}
\usepackage{epsfig,amsmath}
\usepackage{amssymb}
\usepackage{subfigure}
\usepackage{float}
\usepackage[usenames,dvipsnames]{color}
\usepackage[pagebackref=false, colorlinks=true]{hyperref}
\definecolor{redish}{rgb}{0.7,0.2,0.0}  
\definecolor{bluish}{rgb}{0.2,0.5,0.8}

\hypersetup{linkcolor=redish,          
                  citecolor=blue,        
                  filecolor=magenta,      
                  urlcolor=bluish}          

\DeclareFontFamily{U}{rsfs}{}         
\DeclareFontShape{U}{rsfs}{m}{n}{<5> rsfs5 <6><7> rsfs7          %
  <8><9><10><10.95><12><14.4><17.28><20.74><24.88> rsfs10}{}     %
\DeclareMathAlphabet{\mathfs}{U}{rsfs}{m}{n}

\def \O{\Omega}

\def \f{\frac}
\def \w{\wedge}
\def \o{\omega}

\def \a{\alpha}

\def \t{\tilde}
\def \O{\Omega}
\def \g{\gamma}

\def \e{\epsilon}
\def \p{\partial}

\def \a{\alpha}

\begin{document}

\title{Spinning Gyroscope in an  Acoustic Black Hole :
Precession Effects and Observational Aspects}

\author{Chandrachur Chakraborty}
\email{chandra@pku.edu.cn} 
\affiliation{Kavli Institute for Astronomy and Astrophysics, Peking University, Beijing 100871, China}

\author{Parthasarathi Majumdar}
\email{bhpartha@gmail.com}
\affiliation{Indian Association for the Cultivation of Science, Jadavpur, Kolkata 700032, India}

\begin{abstract}
The exact precession frequency of a freely-precessing test gyroscope is derived for
a $2+1$ dimensional rotating acoustic black hole analogue spacetime, without making 
the somewhat unrealistic assumption that the gyroscope is {\it static}. We show that, as a consequence, the gyroscope
crosses the acoustic ergosphere of the black hole with a {\it finite} precession 
frequency, provided its angular velocity lies within a particular range determined 
by the stipulation that the Killing vector is timelike over the ergoregion. 
Specializing to the `Draining Sink' acoustic black hole, the precession frequency is
shown to {\it diverge} near the acoustic {\it horizon}, instead of the vicinity of the
ergosphere. In the limit of an infinitesimally small rotation of the acoustic black 
hole, the gyroscope still precesses with a {\it finite} frequency, thus confirming a
behaviour analogous to {\it geodetic} precession in a physical non-rotating spacetime
like a Schwarzschild black hole. Possible experimental approaches to detect acoustic
spin precession and measure the consequent precession frequency, are discussed.
\end{abstract}

\maketitle

\section{Introduction}

In general relativity, spacetime curvature causes the spin of a freely-precessing
gyroscope travelling along a geodesic to undergo a precession known as `geodetic 
precession' or `de-Sitter (dS) precession', as was first  predicted by Willem de Sitter
in 1916 \cite{chiba,kro}. If the spinning test object (gyroscope) happens to move 
in a stationary axisymmetric spacetime like a rotating black hole, it undergoes an
additional precession known as Lense-Thirring precession arising due to the dragging
of inertial frames by the rotating black hole spacetime. This latter precession
persists even if the trajectory of the gyroscope is not geodesic \cite{str,cp,cmb,cc2,ccb,saj}. 
Therefore, the complete precession frequency of a test gyroscope ought to be computed
taking all these effects into account. In a recent paper \cite{ckp}, the spin 
precession formulation of a test gyroscope, valid in any general stationary and 
axisymmetric spacetime, has been derived. That work was motivated by the need to 
distinguish a superspinar (also known as Kerr naked singularity) \cite{ckj,kjb} from 
a black hole using the spin 
precession of a test gyroscope. The general formalism \cite{ckp} stipulates that 
{\it stationary rotating} gyroscopes can avoid any divergence in their precession 
frequency at the ergosphere. Such a divergence is known to be an artifact of test 
gyroscopes that are assumed to be {\it static} \cite{cm, ckj}.
From a realistic standpoint, an ordinary test object following a timelike trajectory
outside the ergosphere, can scarcely remain static inside the spacelike ergoregion of 
a rotating black hole.

The direct observation of such precession effects is extremely challenging technically,
if not outright impossible, in strong-gravity astrophysical phenomena. Acoustic black
hole analogues offer an alternative option to probe a general relativistic 
astrophysical phenomenon \cite{un, bm} in a comparatively accessible laboratory 
setup \cite{st, tor, liao}. An incipient exploration of the acoustic Lense-Thirring 
precession has been performed recently \cite{cgm} for Draining Sink (DS) vortex flows 
described in terms of rotating acoustic black hole analogues. There, it has been 
shown that the acoustic LT precession frequency increases unboundedly as one 
approaches the ergosphere. This is very likely an artifact, as mentioned, of 
considering test gyroscopes which are {\it static} inside the ergoregion. The 
artifactual aspects of the previous assay are discarded here by considering instead
test gyroscopes which undergo rotation as they move in the acoustic black hole 
geometry. The range of allowed angular velocities of the rotating gyroscope is
determined by requiring that the Killing vector field $K = \p_t + \Omega \p_{\phi}$ 
now remains timelike throughout the ergoregion. In the case of the DS 
acoustic black hole, therefore, one seeks to reconsider the precession of a test 
gyroscope. As we shall show, a gyroscope is seen to cross the ergosphere without 
any spectacular enhancement in its precession frequency. This requirement restricts
possible angular velocities of the test gyroscope to lie within a certain range. 
The enhancement of the precession frequency {\it reappears} however, as we shall
demonstrate, when the gyroscope approaches the acoustic {\it event horizon}, 
corresponding to the Killing vector field $K$ turning null on the acoustic horizon. 

A key question in acoustic analogue gravity work is of course that of experimental 
or observational accessibility of such precession effects. In normal inviscid, 
barotropic fluids, the primary excitations are the phonon perturbations which 
usually are assumed to have no spin polarization. Thus, how does one conceive of 
a spinning test gyroscope using such phonons ? There is apparently no easy answer
to this question. In ref. \cite{cgm}, it has been suggested that if the phonons 
representing acoustic perturbations inside the fluid have an intrinsic spin, in 
addition to their orbital angular
momentum, which is free to precess around the rotation axis of the background flow,
one can study such a `spin precession' as a gyroscopic precession. The notion of an 
intrinsic phonon spin has been proposed by Zhang and Niu \cite{zhang2014} for spin 
relaxation in ionic crystals exposed to uniform magnetic fields, based on the Raman
spin-phonon interaction which is linear in the phonon momentum \cite{ray1967}. An 
elaboration of this notion of phonon spin has also been presented by Garanin and Chudnovsky \cite{garanin2015}. Another possible scenario which can be useful to observe the `spin precession',
involving \textit{spinor} condensates \cite{pethickbook, fetter2009}.
In this paper, we describe another observational scenario which involves a change of
the fluid itself to a {\it biologically active nematic} fluid : bacteria or active 
particles with a rod-like structure swimming through a solvent fluid. Recently, the
first-ever analysis of hydrodynamics of such nematic fluids, from the acoustic analogue
gravity standpoint, has just appeared as an e-print \cite{bkm}. 

We organize the paper as follows : we derive the general spin precession formalism in 
acoustic DS geometry in section \ref{gf}. Section \ref{geo} is devoted to
show that a spin can precess even in the `non-rotating' acoustic analogue spacetime. 
We elaborate on experimental and observational scenarios for
kinematic gravitational precession effects in section \ref{obs}. 
Finally, we conclude in section \ref{con}.

\section{General formalism of spin precession in (2+1)D geometry \label{gf}}~

In a rotating spacetime, an observer can remain still without changing its location with respect to infinity, only outside the ergoregion. Such an observer is called as a static observer and its four-velocity is written as:
$u_{\rm static}^{\a}=u_{\rm static}^0(1,0,0,0)$. In contrast, an observer can hover very close the horizon of a rotating black hole, if the observer rotates around the black hole  with respect to infinity. Such observer is called as a stationary observer and its four-velocity is written as: $u_{\rm stationary}^{\a}=u_{\rm stationary}^0(1,0,0,\O)$ \cite{ckp}, where  $\O$ is the angular velocity of the observer. The general spin precession frequency of a {\it static} gyro relative to a Copernican frame \cite{str, cgm} was derived for a four dimensional stationary spacetime in \cite{str}, and it has recently been extended for a {\it stationary} gyro in \cite{ckp}.

Let us now consider a test gyroscope, which moves along a Killing trajectory in a stationary $(2+1)$D spacetime. The spin of such a test spin undergoes Fermi-Walker 
transport along the 4-velocity of the test body \cite{str}
\begin{equation}
u=(-K^2)^{-\f{1}{2}} K,
\end{equation}
where $K$ is the timelike Killing vector field appropriate to stationarity of the spacetime. In this special situation, 
it is known that the gyroscope precession frequency coincides with the 
vorticity field associated with the Killing congruence, i.e., the gyro
rotates relative to a corotating frame with an angular velocity.
Referring to previous work \cite{cm, ckp, cgm} for detailed derivations,
the spin precession of a test spin in $(2+1)$ dimensional spacetime can be
expressed as 
\begin{eqnarray}
 \O_{(2+1)}=\f{1}{2K^2}*(\t{K} \w d\t{K})
 \label{b}
\end{eqnarray}
following \cite{str}, where $\O_{(2+1)}$ is the spin precession rate of a test spin relative to a Copernican
frame of reference in coordinate basis,
$\t{K}$ is the dual one-form of $K$ and $*$ represents the Hodge star 
operator or Hodge dual. By spin we mean either the polarization vector of a
particle (i.e., the expectation value of the spin operator for a particle 
in a particular quantum mechanical state) or the intrinsic angular momentum 
of a rigid body, such as a gyroscope \cite{str}. We also recapitulate here that the spin precession frequency $\O_{(2+1)}$ in $(2+1)$D acoustic spacetime is now
a spatial scalar \cite{cgm}. In any stationary spacetime,
$K$ can be expressed as $K=\p_0$ \cite{str} for which Eq.(\ref{b}) reduces to \cite{cgm},
\begin{eqnarray}
\O_{(2+1)}&=&\f{1}{2\sqrt {-g}}\e_{ij}g_{00} \left[\f{g_{0i}}{g_{00}}\right]_{,j}.
\label{lt2d}
\end{eqnarray}
Since the focal point of this paper is to study the spin precession in 
the rotating acoustic DS geometry, we point out here 
that it has two Killing vectors : one is the time translation Killing vector
$\p_0$ and another is the azimuthal Killing vector $\p_{\phi}$. Now, we can construct a new Killing vector from a linear combination, with constant
coefficients, of  $\p_0$ and $\p_{\phi}$. With this motivation, we consider here the precession of the spin of gyroscopes attached to stationary observers, whose velocity vectors are proportional to the
Killing vectors $K=\p_0+\O \p_{\phi}$  \cite{ckp}. These gyroscopes move along the circular orbits around the central object with a constant angular velocity $\O$, which at any given distance ($r$) can be chosen to be in a particular range, so that $K$ is timelike.
However, for a general stationary spacetime which also possesses a spacelike 
Killing vector, we can write down the general timelike Killing vector as :
\begin{eqnarray}
 K=\p_0+\O \p_c ,
 \label{k}
\end{eqnarray}
where $\p_c$ is a spacelike Killing vector in that stationary
spacetime, i.e., the spacetime is isometric vis-a-vis two coordinate directions
: $x^0$ and $x^c$. Therefore, the corresponding co-vector of $K$ can be written as
\begin{eqnarray}
 \t{K}=g_{0\nu}dx^{\nu}+\O g_{\g c}dx^{\g}
\end{eqnarray}
where $\g , \nu=0,c,2$ in 3-dimensional spacetime. 
Separating space and time components we can write 
\begin{eqnarray}
  \t{K}=(g_{00}dx^0+g_{0c}dx^c)+\O (g_{0c}dx^0+g_{cc}dx^c)
  \label{kt}
\end{eqnarray}
and 
\begin{eqnarray}
  d\t{K}=(g_{00,j}dx^j \w dx^0+g_{0c,j}dx^j \w dx^c)
  +\O (g_{0c,j}dx^j \w dx^0+g_{cc,j}dx^j \w dx^c) .
  \label{kto}
\end{eqnarray}
Substituting the expressions of $\t{K}$ and $d\t{K}$ in Eq.(\ref{b}),
we obtain the spin precession frequency in $(2+1)$D spacetime as:
\begin{eqnarray}\nonumber
 \O_{p}&=&\f{\e_{cj}}{2\sqrt {-g}\left(g_{00}+2\O~g_{0c}
 +\O^2~g_{cc}\right)}.
\left[g_{00}^2~\left(\f{g_{0c}}{g_{00}}\right)_{,j}+g_{00}^2~\O\left(\f{g_{cc}}{g_{00}}\right)_{,j}
 + g_{0c}^2~\O^2 \left(\f{g_{cc}}{g_{0c}}\right)_{,j} \right]
 \\
 \label{ltp}
\end{eqnarray}
where we use $*(dx^0 \w dx^j \w dx^c)=\eta^{0jc}
=-\f{1}{\sqrt{-g}}\e_{jc}$ and $K^2=g_{00}+2\O g_{0c}+\O^2 g_{cc}$. 
$\eta^{0jc}$ represent the components of the volume form in $(2+1)$D
spacetime. We note that Eq.(\ref{ltp}) reduces to Eq.(\ref{lt2d})
for $\O=0$, which is only applicable outside the ergoregion.

\subsection{Application to the `Draining Sink' acoustic black hole}

The acoustic analogue of a rotating black hole spacetime is 
best captured by a planar `Draining Sink' flow of an incompressible,
barotropic, inviscid fluid with no global vortex present. The flow is
characterized by the velocity potential 
\begin{eqnarray}
\vec v_b = -\frac{A}{r}~\hat{r}+\frac{B}{r}~\hat{\phi}~ \label{dbvel}
\end{eqnarray}
where $(r, \phi)$ are plane polar coordinates. The two parameters, $A$ (for drain) 
and $B$ (for circulation) are constants, and also analogous to the mass and angular 
momentum of a rotating black hole \cite{tor18}, respectively.
Therefore, the radial component of the fluid velocity is characterized by
$v_r=|A|/r$ and the angular velocity of the flow is described by $\O_a=B/r^2$.
It was shown in Ref.\cite{cgm} that $\O_a$ (or $B$) was responsible for the 
dragging of inertial frames.
It can also be seen from the explicit emerging form of the (2+1)-D
acoustic  black hole metric (Eq.(2) of \cite{cgm} or Eq.(6) of \cite{bm}) ,
\begin{eqnarray}
 ds_{\rm DS}^2=-\left(1-\f{A^2+B^2}{r^2}\right)dt^2+\left(1-\f{A^2}{r^2}\right)^{-1}dr^2
-2B ~d\phi dt+r^2~d\phi^2
\label{met}
\end{eqnarray}
that it describes a `non-rotating' acoustic analogue black
hole geometry for $B=0$, for which we do not see any frame-dragging effect as expected. Another interesting thing is that unlike Kerr black hole,
the ergoregion is spherical in shape in a $(3+1)$D acoustic analogue 
black hole and it does not touch the event horizon in any direction, i.e.,
event horizon and ergoregion both are circular in shape in the $(2+1)$ dimensional `Draining
Sink' geometry. 

Now, using Eq.(\ref{ltp}) we can obtain the spin precession frequency of a
`test spin' in the $(2+1)$D  `Draining Sink' geometry as
\begin{eqnarray}
  \O_{p}= -\f{r_E^2(B-\O r^2)+\O r^2 (r^2-r_E^2)+B\O^2 r^4}
  {r^2[(r^2-r_E^2)+2B \O r^2-\O^2 r^4]} .
  \label{main}
\end{eqnarray}
For $\O=0$, the above Eq.(\ref{main}) reduces to Eq.(9) of Ref.\cite{cgm}.

In the above expression (Eq.\ref{main}), the angular velocity $\O$ of 
the test spin is constrained by the requirement that $K$ remains timelike outside the acoustic horizon at $r=A$.
Therefore, it should satisfy the following condition \cite{ckp}
\begin{eqnarray}
 K^2 = g_{\phi\phi}\O^2+2g_{t\phi}\O+g_{tt} & < & 0,
\end{eqnarray}
i.e., inside the ergoregion the angular velocity can take only those values 
which are in the following range:
\begin{eqnarray}
 \O_- < \O < \O_+ 
\end{eqnarray}
 where, 
\begin{eqnarray}
 \O_{\pm} =\f{-g_{t\phi}\pm \sqrt{g_{t\phi}^2-g_{\phi\phi}g_{tt}}}{g_{\phi\phi}}.
\end{eqnarray}
For this particular metric (Eq.{\ref{met}}), 
\begin{eqnarray}
 \O_{\pm}=\f{1}{r^2}(B \pm \sqrt{r^2-A^2}).
 \label{opm}
\end{eqnarray}
Now, the test spin can take any value of $\O$ between $\O_+$ and $\O_-$.
Here, we introduce a new parameter $q$ to scan the range of allowed values
of $\O$. Therefore, we can write
\begin{eqnarray}
 \O = q~\O_+ + (1-q)~\O_- = \f{1}{r^2}\left[(2q-1)~\sqrt{r^2-A^2}+B \right]
 \label{q}
\end{eqnarray}
where $0 < q < 1$. 

\begin{figure}
\begin{center}
\includegraphics[width=4in,angle=0]{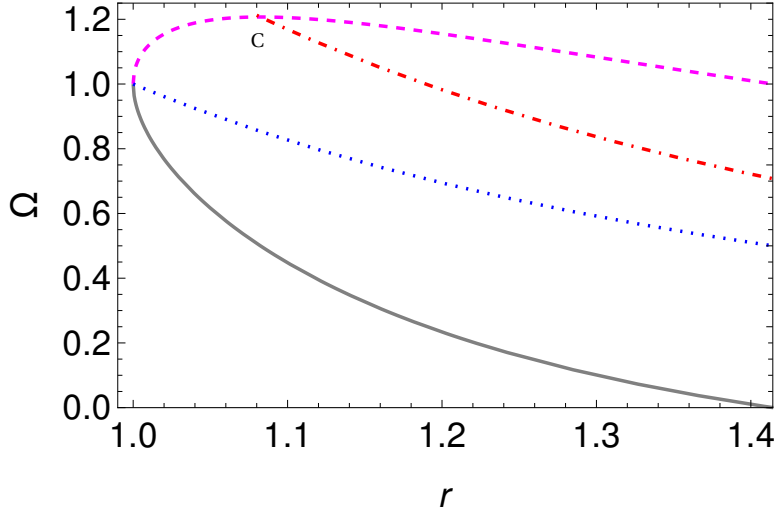}
\caption{\label{fo} Variation  of $\O$ (unit of `length$^{-1}$') versus $r$ 
(unit of `length') for $A=B=1$. $\O$ can take values only in the range: 
$(\O_-,\O_+)$. $\O_+$ and $\O_-$ (unit of `length$^{-1}$') are in dashed
magenta and solid gray respectively and are plotted specifically inside the 
ergoregion, as a function of $r$. It is seen that $\O_{\pm}$ meet at the horizon.
The dotted blue line stands for the frequency of ZAMO (Zero Angular Momentum 
Observer), i.e., $q=0.5$. The orbital frequency, i.e., $\O_{\phi}=r_E/r^2$ 
(see Eq.\ref{op}) which is indicated by the dot-dashed 
red curve, crossed over the $\O_+$ curve at point `C'. It signifies that
the observer moving with $\O_{\phi}$ cannot continue its motion
after reaching at the orbit $r=r_{C}$ which corresponds to the particular
point `C'.}
\end{center}
\end{figure}

Fig.\ref{fo} shows that the test spin can take any value of $\O$ which is fallen between the dashed magenta and the 
solid gray curves. One intriguing feature is that the orbital frequency $\O_{\phi}$
(see Eq.\ref{op} of Appendix \ref{ff}) becomes equal to $\O_+$ at point `C' \cite{cbgm2}. The 
corresponding radius of the orbit ($r=r_{C}$) is
\begin{eqnarray}
r_{C}=\sqrt{2 r_E (r_E-B)}.
\label{rc}
\end{eqnarray}
Therefore, a test spin with $\O_{\phi}$ is unable to continue its motion at $r \leq r_{C}$
 whereas it is possible for $r > r_{C}$. As we have already
mentioned, one can note that the test spin can easily continue its motion with 
any value of $\O$ within the range : $\O_- < \O < \O_+$ in the region $r > r_H$. Fig.\ref{fo} also 
reveals that the $\O_{\pm}$ meet at the horizon with the frequency
\begin{eqnarray}
 \O_H = \f{B}{r_H^2}.
 \label{oh}
\end{eqnarray}
Substituting the expression of $\O$ (Eq.\ref{q}) in Eq.(\ref{main}) we obtain
\begin{eqnarray}
  \O_{p}= -\f{2B(1-2q+2q^2)\sqrt{r^2-A^2}-(1-2q)(r^2-2A^2)}
  {4q(1-q)r^2 \sqrt{r^2-A^2}}  ~,
  \label{main2}
\end{eqnarray}
demonstrating that the spin precession frequency ($\O_p$) becomes arbitrarily large
as it approaches to the horizon ($r \rightarrow A$) for all values of $q$ except 
$q=0.5$. One should note that Eq.(\ref{main2}) is not valid on the horizon 
($r = r_H$) as $K$ (Eq.\ref{k}) must turn null on it.

\begin{figure}
\begin{center}
\includegraphics[width=5in,angle=0]{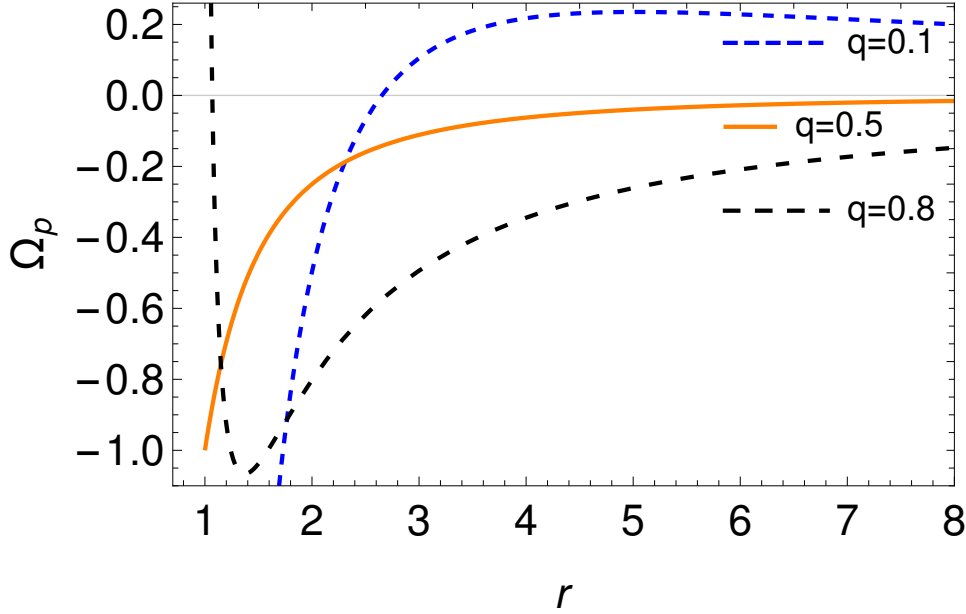}
\caption{\label{fg} Variation  of $\O_p$ (unit of `length$^{-1}$')
versus $r$ (unit of `length') for $A=B=1$. It shows that the spin precession
frequency $\O_p$ becomes arbitrarily large as it approaches to the horizon for any 
value of $q$ whereas $\O_p$ is finite for $q=0.5$ at $r \rightarrow r_H$.}
\end{center}
\end{figure}

We can see the evolution of the spin precession frequency ($\O_p$) of a test spin 
from Fig.\ref{fg} which is plotted for three consecutive values of $q : 0.1,~ 0.5$ 
and $0.8$. We have taken $A=B=1$ (in the unit of length) which means that the
radius of event horizon $r_H=A=1$ and radius of the ergoregion $r_E=\sqrt{2}$.
It can be seen from the figure that $\O_{p}$ vanishes for 
a particular value of $r=r_0$ except for $q=0.5$. The value of $r_0$ for these
cases can be calculated using 
Eq.(\ref{main2}) and setting $\O_p|_{r=r_0}=0$, which comes out as
\begin{eqnarray}
 r_0|_{(0 < q < 0.5)}&=&\f{\sqrt{2Y}}{1-2q}~\left[Y + B(1-2q+2q^2)\right]^{\f{1}{2}}
 \label{rp}
 \\
 {\rm and} \nonumber
 \\
 r_0|_{(0.5 < q < 1)}&=&\f{\sqrt{2Y}}{1-2q}~\left[Y - B(1-2q+2q^2)\right]^{\f{1}{2}} 
 \label{rn}
\end{eqnarray}
where, $Y=\left[r_E^2~(1-2q)^2+4 q^2 B^2~(1-q)^2 \right]^{\f{1}{2}}$. Eq.(\ref{rp})
is valid for $0 < q < 0.5$ and Eq.(\ref{rn}) is valid for $0.5 < q < 1$. 
This means that even the spacetime possesses a non-zero angular velocity
($\O_a \neq 0$ or $B \neq 0$), the test spin does not precesses
at a particular orbit of radius $r_0$. The dashed blue curve of Fig. \ref{fg} shows that the 
spin precession frequency ($\O_p$) first increases with decreasing of $r$, then it becomes 
maximum (at $r=r_p$) and decreases to zero at $r=r_0$ for $q=0.1$.
Surprisingly, it becomes `negative' in the region $r_H < r < r_0$, which means 
that the spin precesses in the reverse direction after crossing the $r=r_0$ orbit. The same feature could be seen
for all values of $q$ within the range
$0 < q < 1/2$. Moreover, as $q$ increases, $r_0$ shifts in the outward direction
and the region ($r_H < r < r_0$) of the `negative precession frequency' becomes 
broader and broader. For $q \rightarrow 0^+$, $r_0 \rightarrow \sqrt{2 r_E~ (r_E+B)}$
which is always greater than $r_H=A$ which means that it is always possible to 
get `negative precession frequency' region for any value of $q ~:~0 < q < 1/2$.

Now, if we consider the value of $q$ as $0.8$, the precession frequency 
curve will show a similar feature as occurs for the first case (i.e., 
$0 < q < 0.5$). The only difference is that the spin precesses in the opposite
direction comparing to the $q < 0.5$ cases and it is evident from the dashed 
black curve of the figure. We point out that the similar trend of the plot could be found 
for all values of $q : 0.5 < q < 1$, i.e., the precession frequency increases 
with decreasing of $r$, becomes maximum and 
then decreases to zero at $r=r_0$ (the value of $r_0$ can be calculated using 
Eq.(\ref{rn}) in this case). As $q$ increases, the value of $r_0$ shifts in the 
outward direction but the maximum value of $r_0$ will be
$r_0 \rightarrow \sqrt{2 r_E~ (r_E-B)}$ for $q \rightarrow 1^-$. In this case, 
the value of $r_0$ is always greater than $r_H$. Therefore, we should get 
a `positive precession frequency' region ($r_H < r < r_0$) for  any value 
of $q~:~ 0.5 < q < 1$.

We note that $q=0.5$ (see the dotted blue curve of Fig.\ref{fo}) is a very special case, as $\O$ (see Eq.\ref{q}) reduces to:
\begin{eqnarray}
\O|_{q=0.5}=\o =-\f{g_{t\phi}}{g_{\phi\phi}}=\f{B}{r^2} = \O_a.
\label{ch}
\end{eqnarray}
In this special case, the above expression gives a particular  angular frequency $(\o)$ of the stationary observer computed in the Copernican frame, which is, in fact, the characteristic ZAMO (Zero Angular Momentum Observer) frequency (see the dotted blue line in FIG. 1).
In this case, test spins attached to stationary observers regard both $+\phi$ 
and $-\phi$ directions equivalently, in terms of the local geometry, and 
{\it see phonons symmetrically} \cite{mtw}. These gyros are non-rotating relative to the 
local spacetime geometry. The angular momentum of such a `locally non-rotating
observer' is zero and is therefore called a zero angular momentum observer (ZAMO),
first introduced by Bardeen \cite{bd,mtw}. Bardeen et al.\cite{bpt} showed that 
the ZAMO frame is a powerful tool in the analysis of physical processes near
astrophysical objects. However, Eq.(\ref{main2}) reduces to 
\begin{eqnarray}
 \O_{p}|_{q=0.5}=-\f{B}{r^2}
 \label{hor}
\end{eqnarray}
for $q=0.5$.
It is clearly seen that Eq.(\ref{hor}) is independent of $A$, which implies that  
the spin precession does not diverge at the horizon $r_H=A$. It is finite for all
values of $r : 0 < r < \infty$ in principle and the spin precession 
rate just outside the horizon is a constant which can be expressed as 
\begin{eqnarray}
 \O_p|_{(q=0.5, r \rightarrow A)}=-\f{B}{r_H^2} = - \O_H.
 \label{ba}
\end{eqnarray}
Noticeably, it is exactly similar to the expression of $\O_H$ (see Eq.\ref{oh})
but the direction is opposite. 
Here, we should note that Eq.(\ref{hor}) diverges only at the `singularity' $r=0$.
Therefore, in principle, the following relation holds everywhere of 
$r : r_H < r < \infty$ in the draining bathtub spacetime for $q=0.5$ :
\begin{eqnarray}
\O|_{q=0.5}=\O_a= -\O_p= \f{B}{r^2} .
 \label{same}
\end{eqnarray}
One can also conclude that a test spin attached to a ZAMO
can easily approach to the event horizon of a `Draining Bathtub' geometry
without facing any major difficulty. We also
note that the precession frequency ($\O_p$) of a test spin attached to a ZAMO
is same with the angular velocity ($\O_a$) of the background fluid flow in the 
draining bathtub geometry but Eq.(\ref{same}) suggests that their directions
are opposite to each other. It is also evident from the solid orange curve of 
Fig.\ref{fg} that the spin precession frequency 
becomes completely `negative' (compared to the first case) for $q=0.5$.
Remarkably, $r_0$ is absent for $q=0.5$, which means that the spin precession
does not vanish except $r \rightarrow \infty$. For $q \neq 1/2$, spin precession 
frequency shows a divergence feature at $r_H=A$, which is similar to
the Kerr case \cite{ckp}.

\subsection{At the boundary of ergoregion}
Now, it is easily seen from Eq.(\ref{main2}) that the divergence of spin 
precession can be avoided at the boundary of the ergoregion, contrary
to the earlier work reported in Ref.\cite{cgm}. To cross the boundary of 
the ergoregion ($r=r_E$) the test spin has to acquire the angular velocity as, 
\begin{eqnarray}
 \O|_{r=r_E}=\f{2qB}{r_E^2} .
 \label{oere}
\end{eqnarray}
Now, substituting the value of $\O$ (Eq.(\ref{oere})) in Eq.(\ref{main}),
we obtain the spin precession rate at the boundary of the ergoregion
\begin{eqnarray}
   \O_p|_{r=r_E}=-\f{4q^2B^2+r_E^2(1-2q)}{4q(1-q)Br_E^2}.
   \label{olte}
\end{eqnarray}
which is finite for the range $0 < q < 1$ (see also Eq.\ref{q}),
{\it contrary to our previous result} (Eq.(9) of \cite{cgm}).
In a special case, say, for $q=0.5$, the above equation (Eq.(\ref{olte})) reduces to
\begin{eqnarray}
   \O_p|_{(r=r_E, q=0.5)}=-\f{B}{r_E^2}.
   \label{olte12}
\end{eqnarray}
which is same as the angular velocity $(\O_a)$ of the DS flow
(Eq.(13) of \cite{cgm}) at the boundary of ergoregion.

\section{Geodetic precession in the `non-rotating'
acoustic analogue black hole \label{geo}}

It has been shown \cite{ckp, cp} that a test spin undergoes precession even in a non-rotating spacetime, if it rotates with a non-zero
angular velocity $\O$. Therefore, the similar incident can happen
in the non-rotating acoustic spacetime also. For $B \rightarrow 0$, the metric (Eq.\ref{met})
reduces to 
\begin{eqnarray}
ds_{\rm DB}^2|_{B=0}=-\left(1-\f{A^2}{r^2}\right)dt^2+\left(1-\f{A^2}{r^2}\right)^{-1}dr^2
 +r^2~d\phi^2
\label{met2}
\end{eqnarray}
and the flow is characterized by the radial component of the fluid velocity potential
\begin{eqnarray}
\vec v_b|_{B=0} = -\frac{A}{r}~\hat{r}~ \label{dbvel2}.
\end{eqnarray}
Though the fluid flow (see Eq.\ref{met2}) does not exactly mimics the 
Schwarzschild geometry but we are reasonably close to it \cite{visser}.
However, using Eq.(\ref{main2}) for $B \rightarrow 0$, one can easily obtain the spin 
precession frequency in the non-rotating acoustic analogue spacetime (Eq.\ref{met2}).
This turns out as:
\begin{eqnarray}
  \O_p|_{B=0}= -\O ~\f{r^2-2A^2}{r^2-A^2-\O^2 r^4}.
  \label{oo}
\end{eqnarray}
where $\O$ is not necessarily to be a function of $r$, rather it can take
any finite value so that $K$ remains timelike. Now, if the test spin 
moves along the circular geodesic, $\O$ should be the orbital frequency,
i.e., $\O_{\phi}=A/r^2 $ (see Eq.(\ref{op}) of Appendix \ref{ff} for the derivation). 
For this particular angular velocity, Eq.(\ref{oo}) reduces to 
\begin{eqnarray}
 \O_p|_{(B=0, \O=A/r^2)} =- \O= -\f{A}{r^2}.
 \label{gp}
\end{eqnarray}
Eq.(\ref{gp}) gives the precession frequency in the Copernican frame, computed with 
respect to the proper time $\tau$ which is related to the coordinate time $t$ via 
$d\tau=\sqrt{1-\f{2A^2}{r^2}}~dt$. Therefore, the precession frequency
in the coordinate basis should be written as
\begin{eqnarray}
 \O' =-\f{A}{r^2}\sqrt{1-\f{2A^2}{r^2}}.
\end{eqnarray}
Now, the difference of $\O_{\phi}$ and $\O'$ could be written as
\begin{eqnarray}
 \O_{dS} =\O_{\phi}-\O'=\f{A}{r^2}\left(1+\sqrt{1-\f{2A^2}{r^2}}\right).
 \label{ds}
\end{eqnarray}
which is identified as the analogous of de-Sitter/geodetic precession
 \cite{chiba,kro} in the non-rotating Schwarzschild black hole, as mentioned in section IV C of 
Ref. \cite{ckp} and the Introduction of this paper.

\begin{figure}
\begin{center}
\includegraphics[width=5in,angle=0]{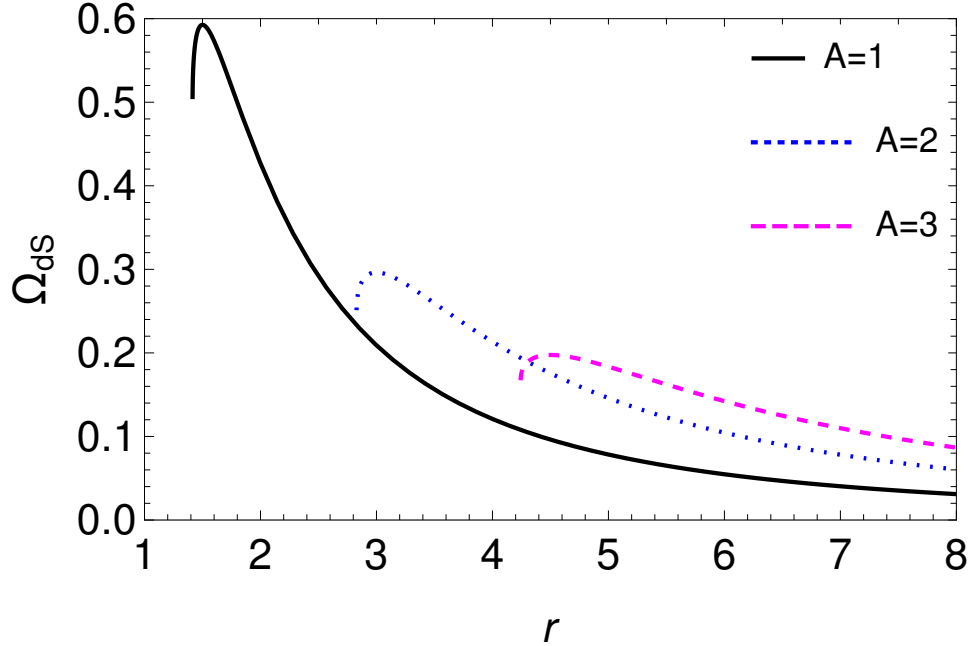}
\caption{\label{fgds} Variation  of geodetic precession $\O_{dS}$ (unit of `length$^{-1}$')
versus $r$ (unit of `length') for the non-rotating acoustic blacks hole with different
$A$ values. It shows that the precession
frequency $\O_{\rm dS}$ initially increases, achieves a peak value and then decreases 
(until it reaches to the orbit of radius  $r=\sqrt{2}A$), as the gyro approaches the 
horizon.}
\end{center}
\end{figure}

One intriguing behaviour that emerges is, the de-Sitter/geodetic precession frequency 
($\O_{\rm dS}$) of the test gyro initially increases, achieves a peak value at
$r=r_{\rm peak}$, then decreases and discontinues for $r < \sqrt{2}A$, as it approaches to the 
horizon (see Fig.\ref{fgds}). The similar
feature cannot occur for Schwarzschild black hole, as the ISCO (innermost
stable circular orbit) is located at $r_{\rm ISCO}=6M$ in this case. On the other hand,
the stable circular orbits exist everywhere (i.e., $r \geq r_H$) in the DS black hole, 
as is pointed out in Appendix \ref{ff} and therefore gyro can approach the horizon
using the stable orbits. However, the gyro does not show geodetic precession 
for $r < \sqrt{2}A$ as Eq.(\ref{ds}) becomes imaginary for those values of $r$.
Eq.(\ref{rc}) also reveals that the orbit of radius $r=\sqrt{2}A$ coincides
with $r_C$ for $B=0$, which is null. Therefore, the test gyro could not be able to
continue its stable `geodesic' motion in those circular orbits which are 
located at : $A < r \leq \sqrt{2}A$, although those orbits are mathematically stable
(see Appendix \ref{ff}). However,
as we have mentioned that the geodetic precession frequency becomes maximum
at $r=r_{\rm peak}$, one can obtain $r_{\rm peak}=1.5A$ differentiating Eq.(\ref{ds})
with respect to $r$ and setting it to zero for $r=r_{\rm peak}$. Thus, the 
maximum geodetic precession frequency achieved by a gyro at $r=1.5A$ would be :

\begin{eqnarray}
\O^{\rm max}_{\rm dS}|_{r=1.5A}=\f{16}{27A} \approx 0.593A^{-1}.
 \end{eqnarray}

\section{Observational Prospects\label{obs}}

The crucial requirement in the fluid with phonon excitations is the existence of an anisotropy which embodies a precessing gyroscope. In general, this is impossible for a phonon fluid, since phonons do not carry any spin. However, as argued by Garanin and Chudnovsky \cite{garanin2015}, circular shear deformations in rotating systems like the DS induces an anisotropy at the classical level in the background fluid. The entire system is of course rotationally invariant, which implies that the shear anisotropy must have a compensation. If the lattice deformation caused by the shear induces a phonon spin through a magnetic Raman spin-phonon interaction, this provides for a mechanism to compensate the shear anisotropy. Recently, Zhang and Niu \cite{zhang2014} have argued that this indeed happens in certain paramagnetic materials, providing the possibility of a phonon spin. The question remains as to whether this magnetic effect can be replicated in condensate systems with atoms in the hydrodynamic approximation.

\begin{figure}
\begin{center}
\includegraphics[width = 7 cm,angle = -90]{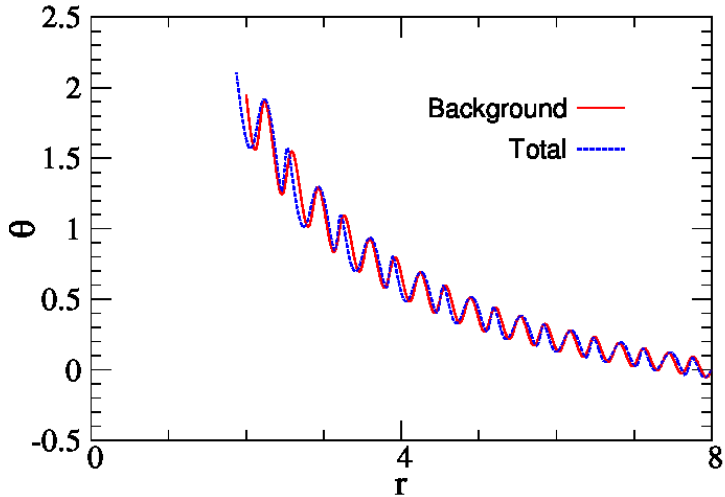}\\
\caption{\label{fg4} Dynamics of the orientation $\theta $ of active particles
with radial coordinate $r$; with $\omega =0.001, A=1, B=10,
\epsilon =0.1$. The initial conditions are $x=8,y=0,\theta =0$.} 
\end{center}
\end{figure}

A very different idea is the possibility of an acoustic analogue black hole in an active nematic fluid with bacteria swimming in it. One can associate with these bacteria an orientation (`polarization') which introduces an intrinsic degree of anisotropy. Typically, this orientation has a time dependence described by \cite{giomi,bkm}
\begin{equation}
 [\partial_t+{\bf v}\cdot\mathbf\nabla]p_i+\omega_{ij}p_j\\[5pt] =\delta_{ij}^T\left[\lambda u_{jk}p_k+\kappa\nabla^{2} p_{j}\right] ~\label{pcont}
\end{equation}
with $\gamma',\kappa,\lambda $ being constants, $\delta_{ij}^{T}=\delta_{ij}-p_{i}p_{j}$ being the
transverse projection operator$~,~u_{ij}  \equiv (\partial_{i}v_{j}+\partial_{j}v_{i})/2~,
{\bf v}=\vec{v_b}+\epsilon \vec{v_1}$ with $\epsilon << 1$, $\omega_{ij} \equiv (\partial_{i}v_{j}-\partial_{j}v_{i})/2$, and $D_{ij}$ being the effective diffusion tensor governing orientation-dependent diffusion of active nematics and is given by
\begin{equation}
D_{ij} = D(\delta_{ij}-\xi p_{i}p_{j}) ~\label{difc}
\end{equation}
where $D$ is the diffusion constant and $\xi$ is another constant related to how much 
diffusion is influenced by the alignment of active particles. The parameter 
$|\lambda|$ measures the alignent of the swimming bacteria to a shear flow in the 
background fluid. The energy lost due to deforming the polarization field of the
particles generated by the aligning interactions between individual particles is 
represented in the last term in right hand side of Eq. (\ref{pcont}). In the limit 
of low background concentration of active particles (bacteria), the active particles 
may serve as freely-precessing gyroscopes, provided one can ignore their 
self-interaction. In this case, the dynamic behaviour of the orientation can be 
graphically represented as in Fig.\ref{fg4} (or, Fig.2(a) of \cite{bkm}). 

It is obvious that there is marked monotonic functional dependence on the radial distance from
the ergosphere. This is precisely what may be quite useful in an accurate measurement of the 
Lense-Thirring frequency without invoking any weak field approximation. While we no longer claim that the precession frequency diverges on the ergosphere, there is still a substantial enhancement outside the ergoregion. This is where laboratory measurements are the most practicable.

\section{Conclusions and discussions\label{con}}
We have derived the exact spin precession frequency in the $(2+1)$D
stationary and axisymmetric spacetime. From this general formulation,
we have shown that the spin precession frequency becomes arbitrarily large
as it approaches to the horizon. We have also shown that a test spin attached with the ZAMO can 
reach close to the horizon of the draining sink geometry without 
facing any major problem, i.e., its precession frequency remains finite.
Contrary to our earlier work \cite{cgm}, it has been shown here that a test spin can cross
the boundary of the ergoregion 
of an acoustic black hole with a finite precession frequency, if the spin 
possesses a non-zero finite angular velocity $\O$ in a particular range. 
We note that the results in Ref.\cite{cgm}
was a special case of this general formalism and valid only for the 
test spin attached to a static observer. 

The notion of the test spin in the $(2+1)$D acoustic analogue spacetime 
has been described in the Introduction but one can question that how 
the `test spin' acquires the various values of $\O$ which has been 
specified in Eq.(\ref{q}). Though the phonon-magnon interactions
with possible spin-dependent coupling to phonons
in spinor condensate are yet to be observed, one can speculate that the
coupling should be different for different ionic crystals, discussed in section 5.1
of \cite{cgm}. This coupling parameter of a particular ionic crystal could
be parameterized with the parameter $q$ of this paper. Moreover, in a very recent
article \cite{bps}, an effective magnetic interaction due to the curvature coupling
of the quasiparticles has been obtained, which could be think as an equivalent
to the `spin-gravity coupling' in the strong gravity regime. One can try to
parametrize the notion of curvature coupling using the parameter $q$.

Finally, the very recent assay \cite{bkm} on discerning an acoustic black hole 
analogue in an active nematic fluid, raises very interesting prospects of an accurate
measurement of the Lense-Thirring precession due to acoustic inertial frame dragging.
If issues regarding viscosity in such fluids can be dealt with by lowering the 
concentration, so that bacteria may indeed propagate with the speed of the fluid,
then experimental viability of this very novel, interdisciplinary approach to
observation of the full effect of `acoustic spin precession' is perhaps the 
best of all methods attempted.

\appendix

\begin{appendix}

\section{\label{ff}Derivation of some important observables in DS black hole}
In section \ref{gf}, we have derived the general spin precession frequency 
of a test spin in the $(2+1)$ DS geometry. We already know that a test spin does 
not move along the geodesic in general \cite{hoj,abk}.
The spin moves along an arbitrary timelike Killing vector field $K$.
In general relativity, the behaviour of a spin 
is different in the strong gravity regime due to the spin-gravity coupling.
This coupling is negligible in the weak gravity regime and that's why one 
can assume that the test spin moves along a geodesic \cite{jh} in the weak gravity 
regime. This is indeed a fair assumption. Therefore, at a reasonably large
distance from the horizon of the draining sink, we can consider 
that the test spin rotates in a circular `geodesic' and we can derive the 
orbital frequency ($\O_{\phi}$) as well as the radial epicyclic frequency ($\O_r$)
experienced by it. From the general expression of the orbital frequency 
\begin{equation}
 \O_{\phi}=\dot{\phi} / \dot{t}= d\phi / dt = 
 \left[-\p_rg_{t\phi}\pm \sqrt{(\p_rg_{t\phi})^2-
 \p_rg_{tt}~\p_rg_{\phi \phi}}\right] / \p_rg_{\phi\phi}
 \end{equation}
we obtain the orbital frequency {\footnote{it is also known as the Kepler
frequency in astrophysics}} in the DS geometry
(see Eq.\ref{met}) as \cite{cgm} 
\begin{eqnarray}
 \O_{\phi}=\f{r_E}{r^2}.
 \label{op}
\end{eqnarray}
In this spacetime, the proper angular momentum ($l$) is written as : 
\begin{eqnarray}
 l=\f{L}{E}&=&-\frac{g_{t\phi}+\Omega_{\phi} g_{\phi\phi}}{g_{tt}+\Omega_{\phi} g_{t\phi}}
\\
&=& \f{r^2~(r_E-B)}{(r^2-r_E^2+Br_E)} .
 \end{eqnarray}
Now, the general expression for calculating the radial ($\O_r$) epicyclic
frequency is \cite{cbgm, cpjcap}
\begin{eqnarray}
 &&\O_r^2=\f{(g_{tt}+\O_{\phi}g_{t\phi})^2}{2~g_{rr}}\left[\p_r^2\left({g_{\phi\phi}}/{Y}\right)
 +2l~\p_r^2\left({g_{t\phi}}/{Y}\right)+l^2~\p_r^2\left({g_{tt}}/{Y}\right) \right]
 \label{re}
\end{eqnarray}
where 
\begin{eqnarray} 
 Y=g_{tt}g_{\phi\phi}-g_{t\phi}^2=A^2-r^2 .
\end{eqnarray}
Using Eq.(\ref{re}) square of the radial epicyclic frequency can be obtained as :
\begin{eqnarray}
 \O_r^2=\f{4r_E^2(r_E-B)^2}{r^6}=\f{4}{r^2}~\O_{\phi}^2~(r_E-B)^2 .
 \label{ref}
\end{eqnarray}
It is well-known to us that the square of the radial epicyclic 
frequency is equal to zero at the innermost stable circular orbit (ISCO) and it is
negative for the smaller radius, which shows the radial instabilities
for orbits with radius smaller than
the ISCO. Interestingly, it can be seen from Eq.(\ref{ref}) that the stable
circular orbits exist everywhere in this draining sink spacetime 
for any value of $r \geq r_{H}$. ``Innermost'' is not applicable here. 
$\O_r^2$ could not be negative for any value of $r$ and thus radial
instability is completely absent in this spacetime.
Now, the periastron precession rate or precession rate of the orbit can be calculated as 
\begin{eqnarray}
 \O_{per}=\O_{\phi}-\O_r=\O_{\phi}\left[1-\f{2}{r}(r_E-B)\right] .
  \label{per}
\end{eqnarray}
Therefore, it may also be possible to see the non-zero precession of the phonon 
orbit in the DS geometry, so long as one restricts observation to the
region outside the horizon.
\end{appendix}
\\

{\bf Acknowledgements :} 
One of us (CC) thanks Oindrila Ganguly for useful discussions on this topic. CC also gratefully acknowledges support from the National Key R\&D Program of China (Grant No. 2016YFA0400703) and the National Natural Science Foundation of China (Grant No. 11750110410).

\end{document}